\documentclass[]{aa}
\usepackage{graphicx}
\usepackage{txfonts}
\begin{document}

\title{Sporadic and intense accretion in a 1 Myr-old brown dwarf candidate}

   \author{D. Nguyen-Thanh
       \inst{1,2,3}
       \and
       N. Phan-Bao
       \inst{1,2}
       \and
       S. J. Murphy
       \inst{4,5}
       \and
       M. S. Bessell
       \inst{5}
       }

\offprints{N.~Phan-Bao} 

\institute{Department of Physics, International University, Ho Chi Minh City, Vietnam. \\
           \email{pbngoc@hcmiu.edu.vn}
           \and
           Vietnam National University, Ho Chi Minh City, Vietnam.
          \and         
          Faculty of Physics and Engineering Physics,
          University of Science, Ho Chi Minh City, Vietnam.
          \and
          School of Science, University of New South Wales Canberra, ACT 2600, Australia.
          \and
          Research School of Astronomy and Astrophysics, Australian National University, 
          Cotter Rd, Weston, ACT 2611, Australia. 
          }

      \date{Received; accepted}

\abstract
    {Studying the accretion process in very low-mass objects has important implications for understanding their formation mechanism. 
     Many nearby late-M dwarfs that have previously been identified in the field are in fact young brown dwarf members of
     nearby young associations. Some of them are still accreting. They are therefore excellent targets for further
     studies of the accretion process in the very low-mass regime at different stages.
}
    {We aim to search for accreting young brown dwarf candidates in a sample of 85 nearby late-M dwarfs.
}
    {Using photometric data from DENIS, 2MASS, and WISE, we constructed the spectral energy distribution
     of the late-M dwarfs based on BT-Settl models to detect infrared excesses. We then searched
     for lithium and H$\alpha$ emission in candidates that exhibit infrared excesses
     to confirm their youth and the presence of accretion. 
}
    {Among the 85 late-M dwarfs, only DENIS-P J1538317$-$103850 (M5.5)
      shows strong infrared excesses in WISE bands.
      The detection of lithium absorption in the M5.5 dwarf and its Gaia trigonometric
      parallax indicate an age of $\sim$1~Myr and a mass of 47~$M_{\rm J}$.
      The H$\alpha$ emission line in the brown dwarf shows significant variability
      that indicates sporadic accretion. This 1 Myr-old brown dwarf also exhibits
      intense accretion bursts with accretion rates of up to $10^{-7.9}$$M_{\odot}$~yr$^{-1}$.
}
    {Our detection of sporadic accretion in one of the youngest brown dwarfs might imply
        that sporadic accretion at early stages could play an important role in the formation of brown dwarfs.
        Very low-mass cores would not be able to  accrete enough material to become stars, and thus they end up
        as brown dwarfs.
        }

\authorrunning{Nguyen-Thanh et al.}
\titlerunning{Sporadic and intense accretion in a 1 Myr-old brown dwarf}

  \maketitle

\section{Introduction}
Very low-mass (VLM) objects, with masses below 0.35~$M_{\odot}$ or spectral types
later than M3$-$M4, are expected to form like low-mass stars (e.g., see \citealt{luhman07} and references therein).
However, physical processes at different stages of their formation such as accretion
are still poorly understood. 
VLM populations in nearby young associations 
at different ages, especially objects with detected circumstellar
disks (e.g., \citealt{reiners09,murphy15,boucher16,sil16,murphy18}),
are important resources for studying
the formation mechanism of VLM objects as well as planets around them.

Typically, new candidate members of young associations have been identified
using proper motions, spatial positions, distances, and color-magnitude diagrams  
(e.g., \citealt{murphy15,luhman18}).
Their membership is then confirmed using spectral diagnostics of youth
such as the Li~I doublet line at 6708~\AA,~  and the Na~I doublet at 8183~\AA~
and 8199~\AA.

Circumstellar disks around young stars are warm enough
to produce detectable infrared (IR) excesses. Therefore, infrared photometry has also been used to identify
new disk-bearing members (e.g., \citealt{aven,sil16,boucher16}).
\citet{aven} searched for circumstellar disks around 85 nearby M dwarfs and found no IR excesses
in any of them.
\citet{boucher16} discovered three late-M dwarfs (M4.5$-$M9) and one L dwarf with candidate circumstellar
(primordial or pre-transitional) disks
from a sample of 1600 objects. They are new candidate members of TW Hya ($\sim$10~Myr), Columba ($\sim$42~Myr)
and Tucana-Horologium ($\sim$45~Myr).
\citet{sil16} searched for new members of nearby young associations using Wide-field
Infrared Survey Explorer (WISE) data among a sample of 1774 objects.
The authors found only an M5 dwarf (WISE J080822.18$-$644357.3, hereinafter WISE0808$-$6443) with significant
excesses at 12 and 22~$\mu$m. The excesses indicate that the M dwarf has a circumstellar disk. 
Based on the spatial coordinates, proper motion and photometric distance,
they concluded that WISE0808$-$6443
is a debris disk-bearing 45 Myr-old candidate member of the Carina association.
\citet{murphy18} measured the velocity width of the H$\alpha$ emission line 
and proved that the M5 dwarf is in fact still accreting.
The IR excesses and accretion activity imply that WISE0808$-$6443
likely hosts a primordial disk.
However, the detection of weak millimeter emission and the lack of detectable CO emission
recently reported in \citet{flaherty} indicate that WISE0808$-$6443 hosts a debris disk.

New members of nearby associations have also been revealed serendipitously by 
spectroscopic observations of nearby late-M dwarf candidates in the field
(e.g., \citealt{reiners09,martin10,pb17}; \citealt{rie19} and references therein).
DENIS-P~J0041353$-$562112, which was initially identified as a nearby M7.5 dwarf \citep{pb01,pb06a}
turns out to be a young brown dwarf (BD) binary (M6.5+M9.0, \citealt{reiners09,reiners10}) member of
the Tucana-Horologium association. Surprisingly, the primary M6.5 dwarf component is still accreting. 
\citet{martin10} spectroscopically observed 78 ultracool dwarf candidates in the field and
revealed seven new members of the Upper Scorpii OB association.
\citet{pb17} found a new member of the $\beta$ Pic
moving group and other young candidate members. 

In this paper, we report our
search for new disk-bearing late-M dwarfs from a sample of 85 nearby late-M dwarfs.
Section 2 presents the sample and our search for IR excesses.
In Sect.~3, we present the results of our search, detections of lithium and
observations of H$\alpha$ emission at different epochs in
DENIS-P~J1538317$-$103850 (hereinafter DENIS1538$-$1038).
We then estimate the mass and age of the source.
We also discuss sporadic accretion in DENIS1538$-$1038 and its implication for BD
formation. Section 4 summarizes our results.

\section{Sample selection and a search for IR excesses}
We selected a sample of 85 nearby late-M dwarfs, which have previously been found, from the Deep Near Infrared
Survey of the Southern Sky (DENIS)
\citep{pb01,pb03,crifo,pb06a,pb06c,pb17}.
The late-M dwarfs were originally detected over 5700 square degrees in the DENIS database
at high Galactic latitude with $|b| \geq 30^{\circ}$ \citep{pb01,pb03}, photometric distances
within 30~pc and the color range $2.0 \leq I-J \leq 3.0$ (M6$-$M8). They were spectroscopically observed
to determine spectral types and distances \citep{crifo,pb06a}. One of them is a triple system \citep{pb06c}.
The sample also included a few very nearby late-M dwarfs within 13~pc \citep{pb17}. Totally, we have 85
late-M dwarfs with IR photometry available in the WISE database.
These late-M dwarfs have spectral types ranging from M5.0 to M9.0.
In our previous paper \citep{pb17},
we searched for the Li~I 6708~\AA~ absorption line in 28 of these 85 late-M dwarfs
and detected lithium in five of them. 
Therefore, we classified these five late-M dwarfs as VLM candidate members of young associations
or the young field.
This implies that the sample could contain more potential young VLM objects, especially
disk-bearing objects.

We therefore constructed
the spectral energy distribution (SED) of all 85 late-M dwarfs to search for IR excesses.
We used photometric data with
$I$-band magnitudes from DENIS, $J$, $H,$ and $K_{\rm S}$-band magnitudes
from the Two Micron All Sky Survey (2MASS)
and mid-infrared $W1$, $W2$, $W3,$ and $W4$-band magnitudes from WISE \citep{cutri14}.
Zero magnitude flux values for DENIS \citep{fouque}, 2MASS \citep{cohen}, and WISE \citep{wright}
were used to convert from magnitudes to fluxes.
BT-Settl model atmospheres \citep{allard13} with solar metallicity and surface gravities 
ranging from $\log g =3.0$ to 5.5 
were used to find the best fits of these models to the observed
photometric data. Trigonometric distances were used if available in Gaia.
We then only selected candidates with observed infrared fluxes more than 3$\sigma$ above the model fluxes 
  (where $\sigma$ is the error on the observed fluxes).
There are 11 candidates in total that match this criterion. These candidates
  were further examined visually in the co-added
images from all four WISE bands\footnote{https://irsa.ipac.caltech.edu/applications/wise/}.
The fluxes of a source measured in $W3$ and $W4$ bands are reliable if the source is clearly detected
in the images of these bands.
We also examined the possibility that $W3$ and $W4$ fluxes are contaminated by background objects
using SERC\footnote{https://archive.stsci.edu/cgi-bin/dss\_plate\_finder},
DENIS\footnote{http://cds.u-strasbg.fr/denis.html} and
2MASS\footnote{https://irsa.ipac.caltech.edu/applications/2MASS/IM/interactive.html}
images. 

\section{Results and discussion}

Among the 85 late-M dwarfs, we only detected IR excesses in DENIS1538$-$1038,
indicating that the source hosts an accretion disk
(see Sect.~3.6 for further details). Our spectroscopic follow-up
revealed the presence of lithium in DENIS1538$-$1038, confirming its youth.
We then used the absolute magnitude versus color diagram
to determine the age and the mass of DENIS1538$-$1038.
To estimate accretion rate, we measured 
the velocity width of the H$\alpha$ line profile 
at 10\% of the peak flux.
We also observed DENIS1538$-$1038 at different epochs to
confirm the sporadicity of the accretion
process in the source.
  
\subsection{IR excesses in DENIS1538$-$1038} 
\begin{figure*}
   \centering
     \includegraphics[width=18.0cm,angle=0]{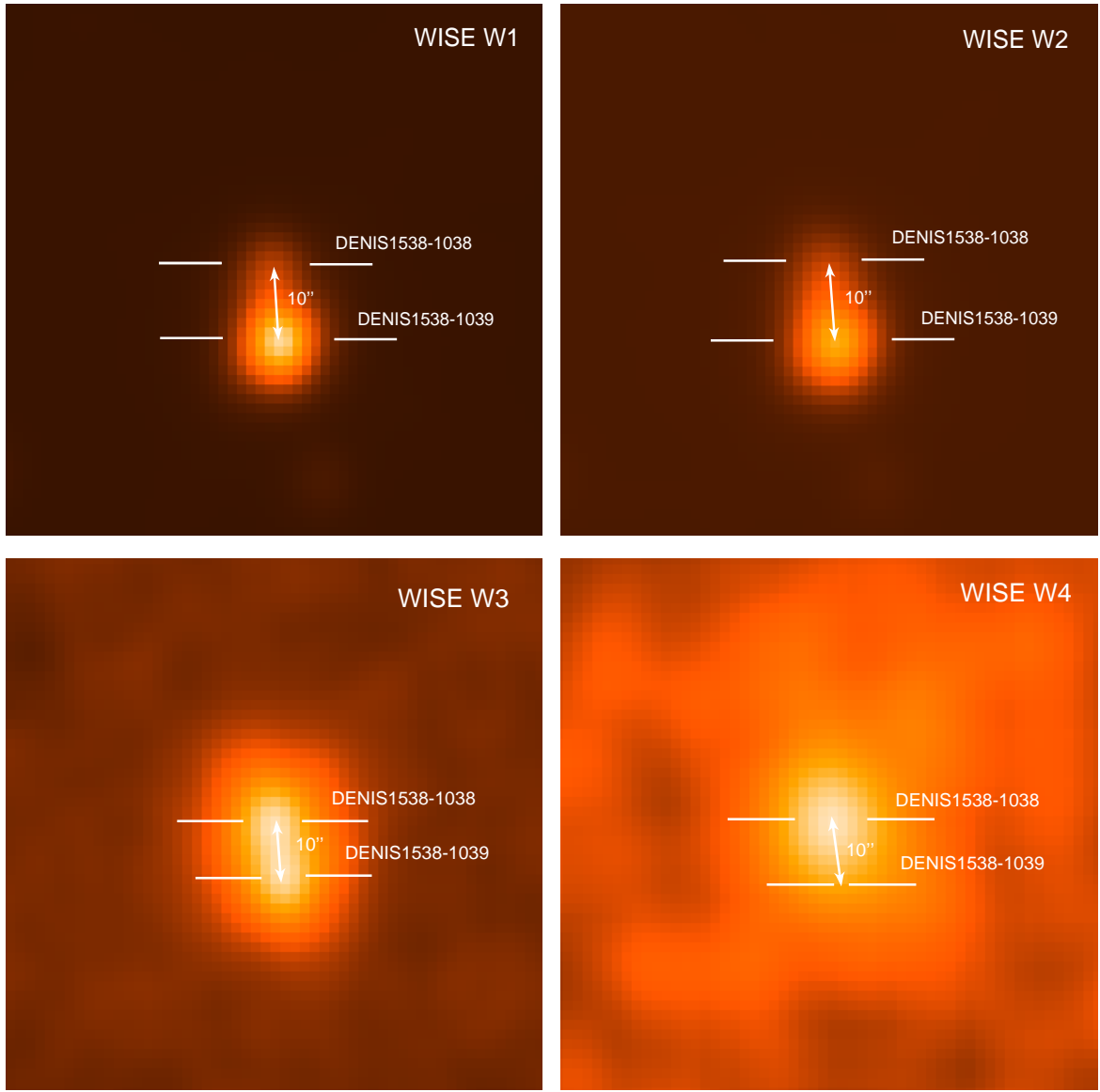} 
     \caption{$W1$, $W2$, $W3,$ and $W4$ WISE images for DENIS1538$-$1038 and DENIS1538$-$1039
       (see Sect.~3.1 for further discussion).
      DENIS1538$-$1039 is not visible in the $W4$ image.
        The bars indicate the positions of DENIS1538$-$1038 and DENIS1538$-$1039.
       A separation of $10''$ between the two objects is also indicated.  
       The charts are $\sim$1.25~$\times$~$\sim$1.25~arcmin$^{2}$.
       North is up, and east is to the left.
}
\label{f1}
\end{figure*}

DENIS1538$-$1038 ($\alpha_{\rm J2000} = 15^{\rm h} 38^{\rm m} 31.70^{\rm s}$, $\delta_{\rm J2000} = -10^{\circ} 38' 50.6''$)
shows significant IR excesses in all four WISE bands (Fig.~\ref{f1}) that are 
above 3, 10, 27 and 9$\sigma$ for bands $W1$, $W2$, $W3,$
and $W4$, respectively. 
\citet{pb03} identified DENIS1538$-$1038 as a low-proper motion
M dwarf ($\mu < 0.1''$~yr$^{-1}$).
DENIS1538$-$1038 has a spectral type of M5.0
and a spectroscopic distance of 31.7~pc \citep{crifo}.    
Figure~\ref{f2} shows the SED of DENIS1538$-$1038.
The SEDs of the 84 remaining M dwarfs are shown in Appendix A.
\begin{figure}
   \centering
     \includegraphics[width=6.0cm,angle=-90]{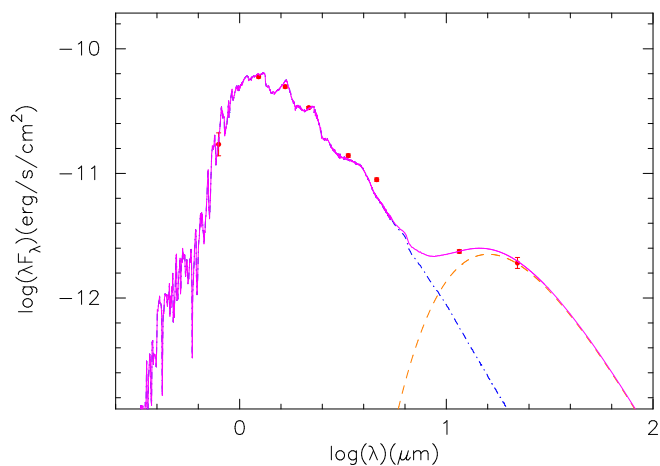}        
     \caption{SED for DENIS1538$-$1038. The blue dash-dotted line
       shows the BT-Settl model atmosphere ($T_{\star} = 2800$~K, $\log g = 3.5$).
       The disk blackbody model ($T_{\rm disk} = 224$~K) is indicated by the orange dashed line.
       Our best fit to the photometric data is shown by the magenta solid line.
}
\label{f2}
\end{figure}
\begin{table*}
  {\small
   \caption{WISE photometry data of DENIS1538$-$1038 and DENIS1538$-$1039}
    \label{tab}
  $$
   \begin{tabular}{lcccccccc}
   \hline 
   \hline
   \noalign{\smallskip}
   Name        & $I_{\rm DENIS}$ & $J_{\rm 2MASS}$ &  $H_{\rm 2MASS}$ & $K_{\rm 2MASS}$ &  $W1$   &  $W2$    & $W3$  & $W4$ \\
               & (mag)          & (mag)        &   (mag)         &  (mag)    & (mag)   & (mag)    &   (mag)  &  (mag)  \\
         \noalign{\smallskip}
\hline
D*1538$-$1038  & 14.36$\pm$0.23 & 12.029$\pm$0.026 & 11.447$\pm$0.030 & 11.096$\pm$0.023 & 10.738$\pm$0.034   & 10.240$\pm$0.034   & 8.758$\pm$0.035 & 6.924$\pm$0.110 \\
D*1538$-$1039  & 11.63$\pm$0.21 & 10.007$\pm$0.024 & ~\,\,9.349$\pm$0.026 & ~\,\,9.111$\pm$0.023 & ~\,\,8.992$\pm$0.022 & ~\,\,8.874$\pm$0.020 & 8.759$\pm$0.031 & $>$8.101      \\
      \noalign{\smallskip}
\hline
    \noalign{\smallskip}
    \hline 
   \end{tabular}
  $$
  \begin{list}{}{}
  \item[] D*: DENIS 
  \end{list}
  }
\end{table*}

\begin{figure}
   \centering
    \includegraphics[width=6cm,angle=-90]{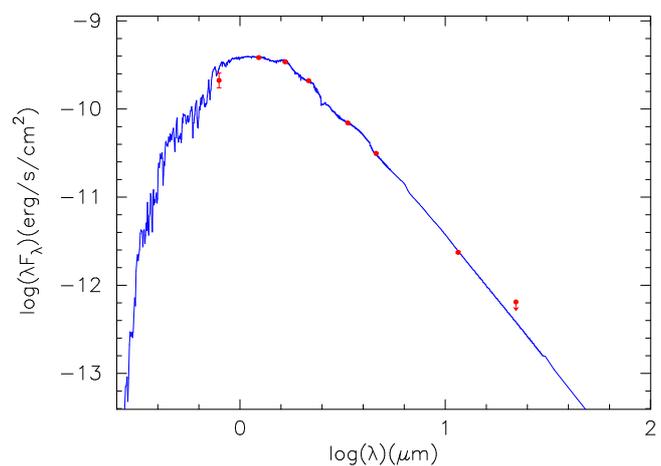}
    \caption{SED for DENIS1538$-$1039. The blue solid line
       shows the BT-Settl model atmosphere.}
\label{f2s}
\end{figure}

During the course of examining the WISE images of DENIS1538$-$1038,
we serendipitously identified DENIS-P~J1538316$-$103900
($\alpha_{\rm J2000} = 15^{\rm h} 38^{\rm m} 31.62^{\rm s}$,
$\delta_{\rm J2000} = -10^{\circ} 39' 00.4''$, hereinafter DENIS1538$-$1039)
at a separation of $\sim$10$''$ from
DENIS1538$-$1038.
Both objects have proper motions and trigonometric parallaxes measured by Gaia.
DENIS1538$-$1038 has a low-proper motion
as measured in \citet{pb03} with $\mu$RA~$= -8\pm9$~mas~yr$^{-1}$
and $\mu$DE~$= -18\pm9$~mas~yr$^{-1}$.
These proper motion values are in agreement with the Gaia measurements,
$\mu$RA~$ = -16.728\pm0.732$~mas~yr$^{-1}$,
and $\mu$DE~$= -22.267\pm0.506$~mas~yr$^{-1}$.
For DENIS1538$-$1039, the dwarf also has a low-proper motion from Gaia
with  $\mu$RA~$= -14.622\pm0.598$~mas~yr$^{-1}$
and $\mu$DE~$= -19.854\pm0.415$~mas~yr$^{-1}$. 
The Gaia trigonometric pallaxes of DENIS1538$-$1038 and DENIS1538$-$1039
are 9.2849$\pm$0.3494~mas (or 107.7$\pm$4.1~pc) and 7.8228$\pm$0.2936~mas (or 127.8$\pm$4.8~pc),
respectively.
Their proper motions and distances suggest that they might belong to
a nearby young association.
A DENIS color $I-J=1.5$ (Table~\ref{tab}) suggests that DENIS1538$-$1039 has a spectral type of $\sim$M4
\citep{bessell91}. We also searched for IR excesses in DENIS1538$-$1039,
however, its SED shows no evidence of a circumstellar disk (Fig.~\ref{f2s}).

We note that, with a separation
of 10$''$,
DENIS1538$-$1038 and DENIS1538$-$1039
are spatially resolved in the WISE $W1$, $W2,$ and $W3$ images (see Fig.~\ref{f1})
with corresponding angular resolutions of 6.1$''$, 6.4$''$, and 6.5$''$ \citep{wright}.
However, the resolution of the $W4$ image of 12$''$ is not capable of resolving
these two sources. DENIS1538$-$1039 could thus contribute its flux
to the IR excess in the $W4$ band of DENIS1538$-$1038 (see Sect.~3.6 for further discussion).
In order to confirm the presence of accretion and the young nature of these two M dwarfs,
we observed the H$\alpha$ emission line and searched for Li~I
absorption at 6708~\AA.

\subsection{Spectroscopic follow-up of DENIS1538$-$1038 and DENIS1538$-$1039}
We observed the two M dwarfs in Apr, May, and Aug 2017, and Jul 2018,
with the Wide Field Spectrograph (WiFeS, \citealt{dopita}) on
the ANU 2.3-m telescope at Siding Spring Observatory. We used the R7000 grating
over a wavelength range of  5280$-$7050~\AA, providing a spectral resolution of $R\approx7000$.
The signal-to-noise ratios of spectra are in the range of 10$-$21.
Table~\ref{tab1} lists the observing dates of the two sources.
We additionally observed them on Mar 22, 2017 with the R3000 grating,
$R\approx3000$ over 5660$-$9640~\AA,~
to estimate the spectral type of the newly identified M dwarf DENIS1538$-$1039.

We then used FIGARO \citep{shortridge} for data reduction.
Only the R7000 spectra were flux-calibrated.
Each night, two flux standards, such as EG~21, EG~131, LTT4364, and L745-46A
were observed.
The R7000 spectra were not corrected for telluric absorption.
For the R3000 spectra, smooth spectrum stars were additionally observed
to remove the telluric lines (see \citealt{bessell99}). 
A NeAr arc was used for wavelength calibration.

\begin{figure}
   \centering
    \includegraphics[width=10cm,angle=-90]{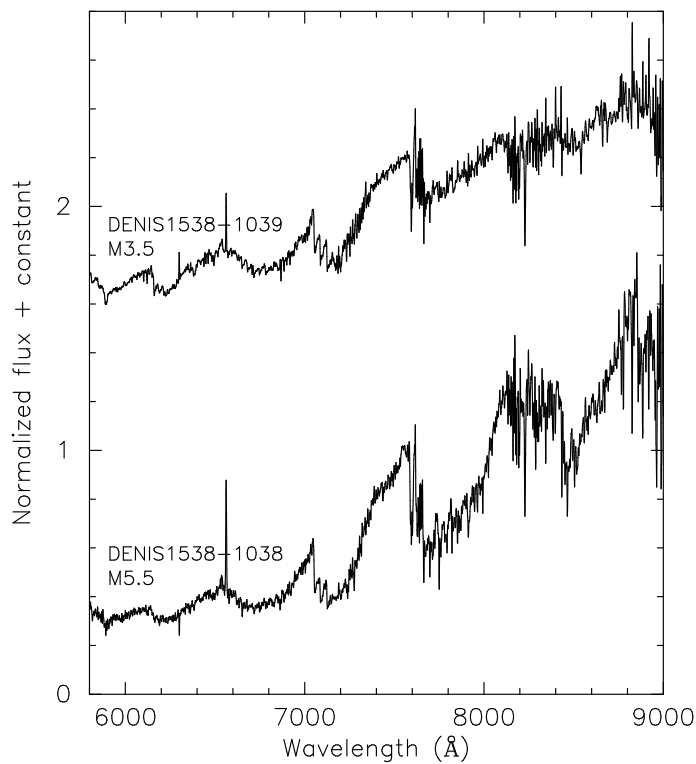}
    \caption{R3000 spectra of  DENIS1538$-$1039 and DENIS1538$-$1038. Spectral types
      adopted from spectral indices VOa, TiO5 and PC3 are also shown.}
\label{f3}
\end{figure}

Figure~\ref{f3} shows the spectra of DENIS1538$-$1039 and DENIS1538$-$1038.
We estimated the spectral type of DENIS1538$-$1039 as in \citet{pb17}.
The adopted spectral type is an average value 
calculated from three indices VOa, TiO5 and PC3.
Our calculation gives a spectral type of M3.5$\pm$0.5 for DENIS1538$-$1039.
With a new R3000 spectrum, we also obtained a spectral type of M5.5$\pm$0.5 for DENIS1538$-$1038,
which is in agreement with the spectral type M5.0$\pm$0.5
determined in \citet{crifo}. In this paper, we adopt the spectral type of M5.5.
We note that the DENIS colors of DENIS1538$-$1039 ($I-J=1.5$)
and DENIS1538$-$1038 ($I-J=2.18$, \citealt{pb03})
are also consistent with the intrinsic colors of young (5$-$30 Myr) M3.5 ($I-J \sim 1.5$)
and M5.5 ($I-J \sim 2.1$) dwarfs (e.g., \citealt{pecaut}).
This implies that the reddening effect is not significant for the two M dwarfs. 
Therefore, we do not consider the de-reddening of SEDs in this paper. 

We used the IRAF task splot to measure equivalent widths (EWs) of the Li~I, H$\alpha$, and He~I at 5876~\AA~ and
6678~\AA, Na~I D1 (5895.9~\AA) D2 (5889.9~\AA), and [O I] 6300~\AA~ lines. The EW uncertainties
were estimated by measuring EWs with different possible continuum
levels and the noise around the regions of interest. In the cases of nondetection of the lines,
upper limits were given by measuring the noise. Table~\ref{tab1} lists our measurements. 

\subsection{Lithium detection}

\begin{figure}
   \centering
    \includegraphics[width=10cm,angle=-90]{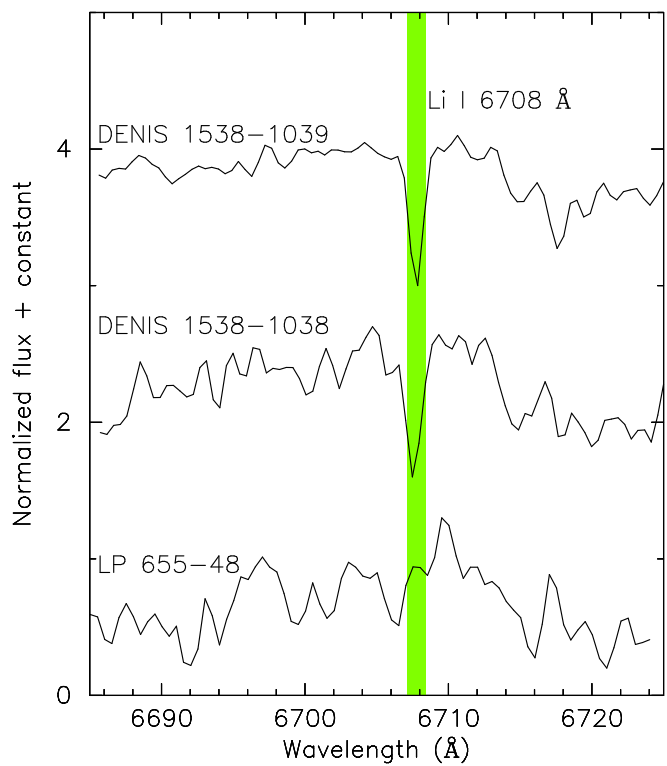}
    \caption{R7000 spectra of DENIS1538$-$1039 and DENIS1538$-$1038
      observed on Apr 19, 2017.
      The spectrum of LP~655-48 with no lithium detection \citep{pb17} is also plotted.
      The region of the Li~I line \citep{pav} is indicated.
}
\label{f4}
\end{figure}

The Li~I doublet line at 6708~\AA~ is clearly detected in both DENIS1538$-$1038
and DENIS1538$-$1039 (Fig.~\ref{f4}).
These detections 
confirm that the two M dwarfs are young (see \citealt{basri00} and references therein).
Our EW measurements of the gravity-sensitive indicator
~Na~I (8183/8199~\AA) as described in \citet{martin10}
from the R3000 spectra
are 4.6$\pm$0.9~\AA~ and 4.6$\pm$1.9~\AA~ for DENIS1538$-$1038 and DENIS1538$-$1039, respectively.
The Na~I EW of DENIS1538$-$1038 is in good agreement with our previous measurement of 4.3$\pm$0.9 \citep{pb17}.
We note that we did not detect lithium in DENIS1538$-$1038 in the previous observation \citep{pb17}
due to a low signal-to-noise ratio.
Here, the signal-to-noise ratio in the lithium region of the spectra is about 15,
significantly higher than the previous one of $\sim$7 \citep{pb17}.
The lithium line is thus revealed in all spectra (Table~\ref{tab1}).

\subsection{Age and mass determination}
\begin{figure}
   \centering
    \includegraphics[width=8cm,angle=-90]{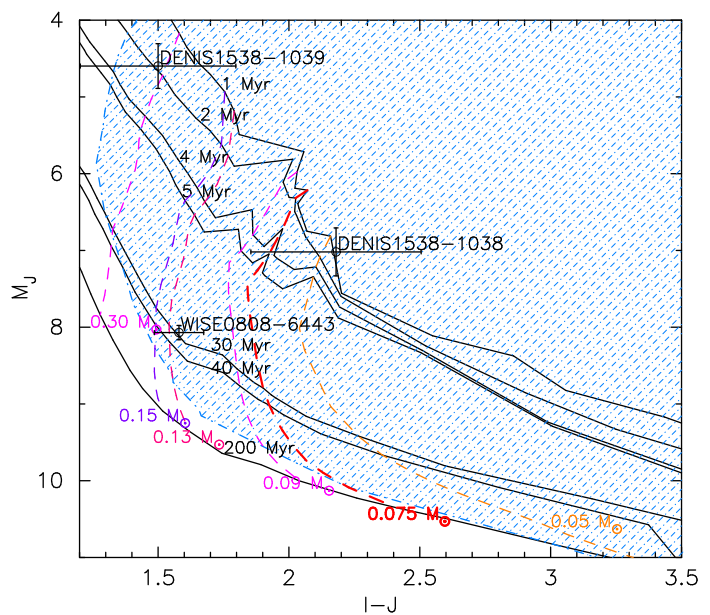}
    \caption{$J$-band absolute magnitude versus color $I-J$ diagram for the accreting M5.5 dwarf (DENIS1538$-$1038)
      and the nonaccreting M3.5 one (DENIS1538$-$1039). Isochrones and mass tracks from the BT-Settl models \citep{allard13}
      are shown. The blue hatched area indicates where lithium remains unburnt and detectable in the stellar photosphere.
      The 45 Myr-old accreting M5 dwarf WISE0808$-$6443
      \citep{sil16,murphy18} is also plotted. With a Gaia parallax of 9.8599$\pm$0.0551~mas, the age of WISE0808$-$6443
      is $\sim$30~Myr and its mass is $\sim$0.12~$M_{\odot}$.
}
\label{f5}
\end{figure}

Using the Gaia distance of 107.7$\pm$4.1~pc and the DENIS $J$-band apparent magnitude $J=12.18\pm0.23$,
we derived an absolute magnitude $M_{\rm J} = 7.02\pm0.31$ for DENIS1538$-$1038.
According to the BT-Settl atmosphere models for the DENIS photometric system \citep{allard13},
the $J$-band absolute magnitude of 7.02 and
the color $I-J=2.18$ of the M5.5 dwarf correspond
to an age of $\sim$1~Myr and a mass of $\sim$47~$M_{\rm J}$ (Fig.~\ref{f5}).
Our estimated mass indicates that DENIS1538$-$1038 is a young BD.
If the error of the $I-J$ color is taken into account,
we derive upper limits of 5 Myr and 90~$M_{\rm J}$ for
the age and the mass of the source, respectively.

For the case of DENIS1538$-$1039, its Gaia distance of 127.8$\pm$4.8 and magnitude
$J=10.13\pm0.21$ give $M_{\rm J} = 4.60\pm0.29$. The derived $J$-band absolute magnitude
and the color $I-J=1.5$ give an age of $\sim$2~Myr and a mass of $\sim$0.33~$M_{\odot}$.
The error of the $I-J$ color and the presence of lithium (Fig.~\ref{f5})
in the M3.5 dwarf place upper limits of 4~Myr and 0.48~$M_{\odot}$ on its age and mass, respectively.

The detections of lithium in both DENIS1538$-$1038 and DENIS1538$-$1039 
clearly indicate the youth of the two sources. Therefore, they are candidate members
of nearby young associations.
To calculate the membership probability of these two M dwarfs in nearby associations, we used
the Bayesian analysis tool BANYAN~$\Sigma$\footnote{http://www.exoplanetes.umontreal.ca/banyan/banyansigma.php}
(see \citealt{gagne18}). With the parallaxes and proper motions from Gaia, we found that DENIS~1538$-$1038
has a membership probability of 31.4\% for Upper Scorpius and 68.6\% for the young field.
For DENIS1538$-$1039, we found a membership probability of 14.3\% for
Upper Scorpius and 85.7\% for the young field. However, the estimated age ranges of 
1$-$5~Myr for DENIS1538$-$1038, and 1$-$4~Myr for DENIS1538$-$1039,
indicate that 
they are more likely to be members of young associations than young field dwarfs.
BANYAN also returns predicted radial velocities for  Upper Scorpius membership to be
$-$9.7$\pm$3.8 and $-$9.9$\pm$3.8~km~s$^{-1}$ for DENIS1538$-$1038 and DENIS1538$-$1039.
At this point, we also measured radial velocities by cross-correlation against standards as described
in detail in \citet{murphy15a}. Our measured radial velocities of DENIS1538$-$1038 and DENIS1538$-$1039 range from $-4.5$$\pm$1.2 to $-14.9$$\pm$1.6~km~s$^{-1}$ and
from $-$2.2$\pm$0.7 to $-$5.3$\pm$1.4~km~s$^{-1}$, respectively.
The predicted values are in agreement with our measurements, further supporting
the membership of these two M dwarfs within Upper Scorpius.
A detailed analysis of radial velocity measurements is to be given in a forthcoming paper.  
A possible scenario to explain the low Upper Scorpius membership probabilities of
DENIS1538$-$1038 and DENIS1538$-$1039
is that the two M dwarfs are members of a new and very young population of 1$-$5~Myr stars
in Upper Scorpius. 
The previous detection \citep{riz15} of young M dwarfs aged 2$-$5~Myr 
in Upper Scorpius supports our scenario. 

\begin{figure*}
   \centering
    \includegraphics[width=18cm,angle=0]{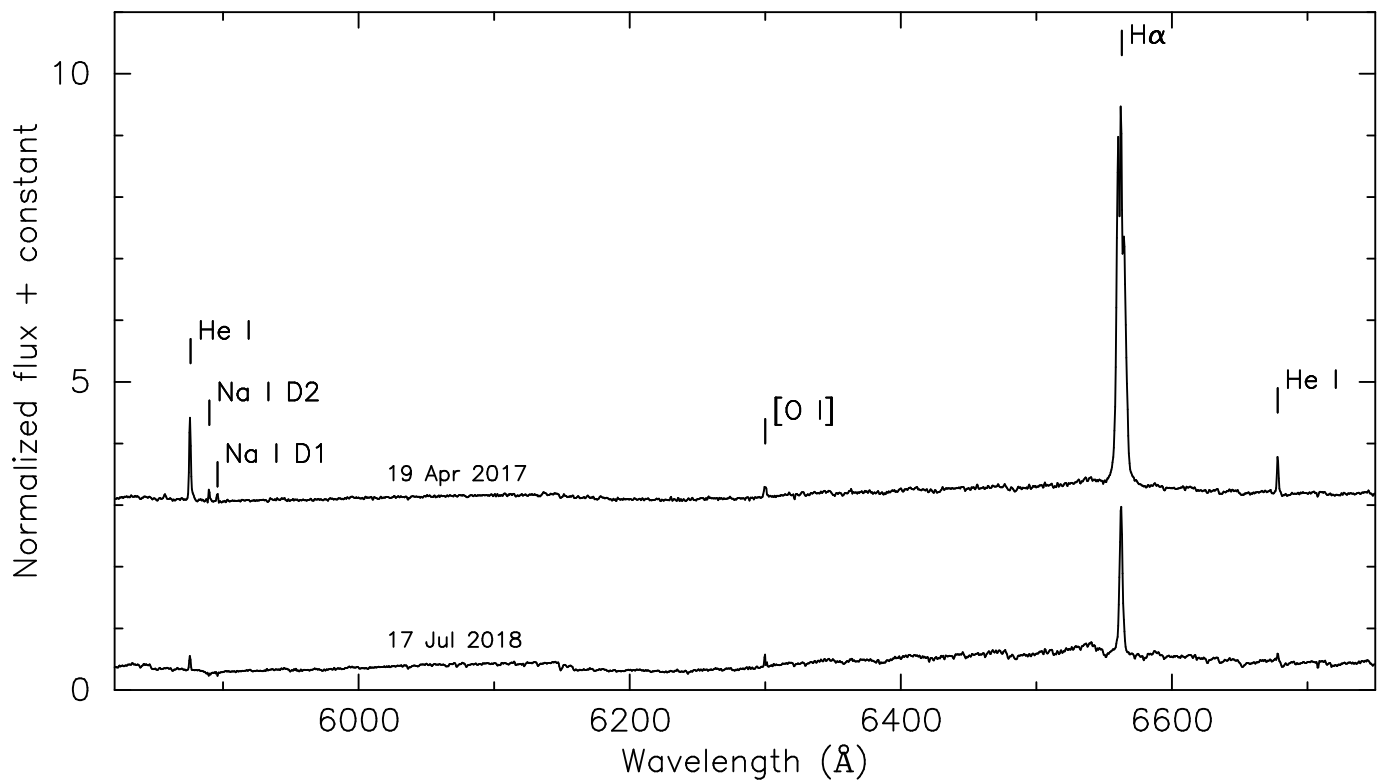}
    \caption{Representative spectra of DENIS1538$-$1038 observed on Apr 19, 2017 and Jul 17, 2018.
      The H$\alpha$, He~I at 5876~\AA~ and 6678~\AA, Na~I D1 (5895.9~\AA) and D2 (5889.9~\AA),
      and [O I]~6300~\AA~ lines are indicated.
      Strong H$\alpha$ emission that is significantly variable indicates sporadic accretion
      (see Sect.~3.5 for further details).
}
\label{f6}
\end{figure*}
\subsection{Sporadic accretion in DENIS1538$-$1038}
\begin{figure}
   \centering
    \includegraphics[width=8cm,angle=0]{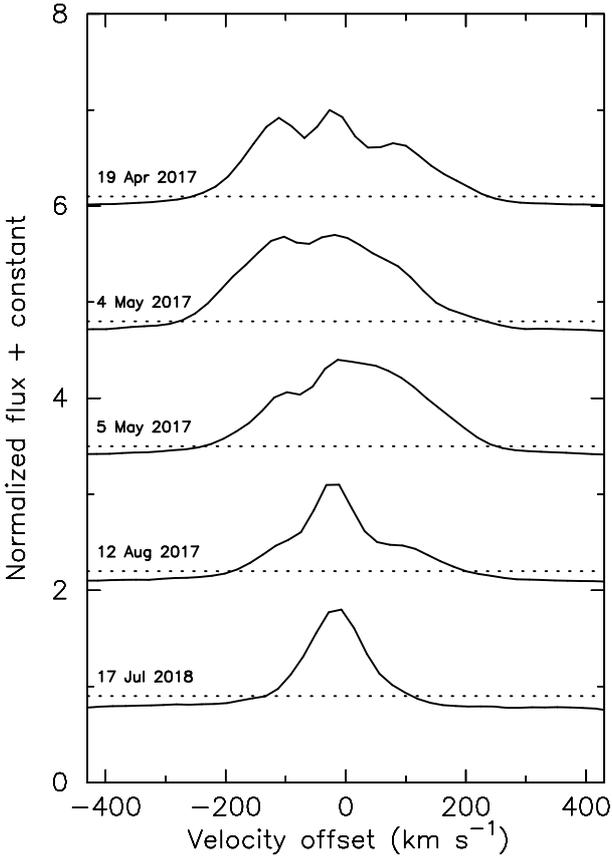}
    \caption{H$\alpha$ velocity profiles of DENIS1538$-$1038. The dashed line indicates the full width at 10\% of the peak.
}
\label{f7}
\end{figure}

Among many indicators of accretion and outflow activities (e.g., see \citealt{herc,rigliaco}),
  we only detected  H$\alpha$, He~I at 5876~\AA~ and
  6678~\AA, Na~I D1 (5895.9~\AA) and D2 (5889.9~\AA), and [O I]~6300~\AA~ in
  the spectra of DENIS1538$-$1038 (Fig.\ref{f6}).
The strong H$\alpha$ emission implies that the observed IR excesses are from
an accretion disk. 
To confirm the presence of accretion,
we measured the velocity width of the H$\alpha$ line profile 
at 10\% of the peak flux (Fig.~\ref{f7}),
which is listed as $v_{10}$ in Table~\ref{tab1}.
\citet{white} proposed $v_{10} > 270$~km~s$^{-1}$
for accreting VLM stars and BDs.  
However, this cutoff can discard VLM objects with very low accretion rates as discussed
in \citet{jay03}.
Consequently, for young VLM stars and BDs, \citet{jay03} proposed $v_{10} > 200$~km~s$^{-1}$
for accretors.
In this paper, we used the \citet{jay03} indicator of accretion.
For the spectra of DENIS1538$-$1038, the $v_{10}$ values all
are above 240~km~s$^{-1}$ (Table~\ref{tab1}).
This confirms that the BD undergoes accretion.
Using the $v_{10}$ versus $\dot{M}$ relation in \citet{natta}, we estimated mass accretion rates
of $10^{-7.9}-10^{-10.5}$~$M_{\odot}$~yr$^{-1}$. The accretion
rates in the M5.5 dwarf observed on 12 Aug 2017 and 17 Jul 2018 are comparable to those measured
in the two M5 dwarfs in Scorpius-Centarus  \citep{murphy15}, but they
are $>$1.4 orders of magnitude more intense on the remaining dates,
implying accretion bursts.

\begin{figure}
   \centering
    \includegraphics[width=8cm,angle=0]{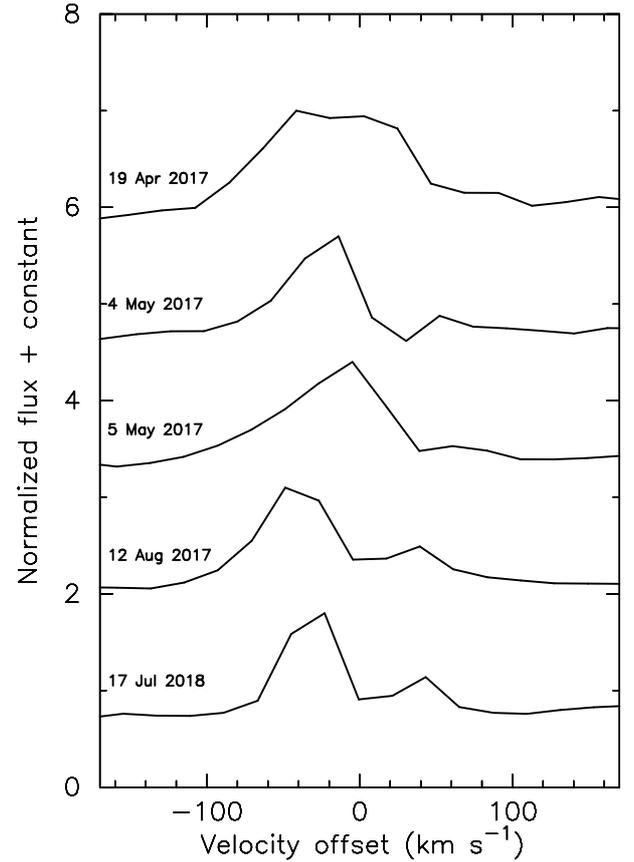}
    \caption{[O I] 6300 \AA~ velocity profiles of DENIS1538$-$1038.
}
\label{f8}
\end{figure}

Here, it is important to discuss the possibilities that
the H$\alpha$ variability observed in DENIS1538$-$1038
could be due to variable accretion (e.g., \citealt{scholz})
or flares as seen in LP~655-48 \citep{pb17}.
Typically,
the [O~I]~6300~\AA~ forbidden emission
is attributed to outflow associated with accretion, while
H$\alpha$ emission can be associated with accretion or with flares.
Therefore, in accretion-driven outflow (e.g., \citealt{cabrit}) 
[O~I]~6300~\AA~ forbidden emission should be correlated with
the H$\alpha$ emission.
In DENIS1538$-$1038,
our EW measurements of the [O~I] emission line (see Fig.~\ref{f8})
as well as Na~I~D1 and D2 (Table~\ref{tab1})
show a slight decrease with decreasing $v_{10}$ of the H$\alpha$ profile.
This suggests that the H$\alpha$ variability
is due to variable accretion, not flaring activity. We therefore conclude that
the accretion in the source is sporadic. 
With an age of $\sim$1~Myr, a mass of $\sim$47~$M_{\rm J,}$ and accretion
rates of $10^{-7.9}-10^{-10.5}$~$M_{\odot}$~yr$^{-1}$,
DENIS1538$-$1038 is one of the youngest M/brown dwarfs yet observed to exhibit sporadic
\citep{lawson,scholz,jay06,murphy11,murphy15,murphy18} and very intense accretion
\citep{com,white,muz03,jay03,natta,muz05,herc,comeron10,manara,sm18}. Sporadic accretion at very early stages with low accretion rates
prevents VLM cores (e.g., \citealt{cuong})
from accreting enough material to become stars \citep{pb14}.

\begin{table*}
  {\footnotesize
   \caption{Observational results of DENIS1538$-$1038 and DENIS1538$-$1039}
    \label{tab1}
  $$
   \begin{tabular}{lccllllllll}
   \hline 
   \hline
   \noalign{\smallskip}
Name              & UT observing &  Li       &  H$\alpha$ & $v_{\rm 10}$(H$\alpha$) & $\log$($\dot{M}$)   &  He I    & He I     & Na I D1 & Na I D2 & [O I]      \\
                  &      date    &           &            &  (km s$^{-1}$)         & ($M_{\odot}$yr$^{-1}$)& 5876 \AA & 6678 \AA & 5895.9 \AA      & 5889.9 \AA      & 6300 \AA   \\
(1)               &    (2)       &  (3)      &   (4)      &  (5)                  &   (6)               &    (7)    & (8)      &  (9)            & (10)           &    (11)     \\
      \noalign{\smallskip}
\hline
D*1538$-$1038     & 19 Apr 2017   &  0.24$\pm$0.02 & ~$-$60$\pm$2     &    495$\pm$6         & ~\,$-$8.1$\pm$0.1   & $-$8.9$\pm$0.6  & $-$2.4$\pm$0.1  & $-$0.8$\pm$0.1  & $-$1.1$\pm$0.1 & $-$1.3$\pm$0.1  \\
                  & 4 May 2017   &  0.38$\pm$0.02 & ~$-$50$\pm$3     &    511$\pm$11        & ~\,$-$7.9$\pm$0.1   & $-$5.8$\pm$0.4  & $-$1.6$\pm$0.1  & $-$0.6$\pm$0.2  & $-$0.7$\pm$0.2 & $-$1.2$\pm$0.1  \\
                  & 5 May 2017   &  0.22$\pm$0.01 & ~$-$49$\pm$1     &    489$\pm$7         & ~\,$-$8.1$\pm$0.1   & $-$7.7$\pm$0.2  & $-$2.9$\pm$0.2  & $-$0.6$\pm$0.1  & $-$0.6$\pm$0.1 & $-$1.2$\pm$0.1  \\
                  & 12 Aug 2017  &  0.47$\pm$0.02 & ~$-$25$\pm$2     &    388$\pm$12        & ~\,$-$9.1$\pm$0.1   & $-$4.1$\pm$0.2  & $-$0.9$\pm$0.2  & ~~~~~$>$$-$0.2  & ~~~~~$>$$-$0.1 & $-$0.8$\pm$0.1  \\
                  & 17 Jul 2018  &  0.41$\pm$0.02 & ~$-$13$\pm$1     &    243$\pm$16        & $-$10.5$\pm$0.2     & $-$2.0$\pm$0.2  & $-$0.3$\pm$0.1  & ~~\,0.4$\pm$0.2     & ~~\,0.9$\pm$0.2    & $-$0.7$\pm$0.1  \\
D*1538$-$1039     & 19 Apr 2017 &  0.48$\pm$0.02 &  $-$1.7$\pm$0.1   &    148$\pm$9        &                     & ~~~~~$>$$-$0.2  & ~~~~~$>$$-$0.1     & ~~\,2.0$\pm$0.2     & ~~\,3.2$\pm$0.2    & ~~~~~$>$$-$0.1  \\
                  & 4 May 2017  &  0.50$\pm$0.02 &  $-$2.3$\pm$0.1   &    149$\pm$7        &                     & ~~~~~$>$$-$0.2  & ~~~~~$>$$-$0.1  & ~~\,1.9$\pm$0.2     & ~~\,2.8$\pm$0.2    & ~~~~~$>$$-$0.1  \\
                  & 5 May 2017  &  0.48$\pm$0.02 &  $-$2.0$\pm$0.2   &    149$\pm$9        &                     & ~~~~~$>$$-$0.2  & ~~~~~$>$$-$0.1  & ~~\,1.9$\pm$0.2     & ~~\,2.9$\pm$0.2    &  ~~~~~$>$$-$0.1  \\
                  & 12 May 2017   &  0.49$\pm$0.02 &  $-$1.9$\pm$0.1   &    157$\pm$12       &                     & ~~~~~$>$$-$0.3  & ~~~~~$>$$-$0.1  & ~~\,2.2$\pm$0.2     & ~~\,2.4$\pm$0.2    & ~~~~~$>$$-$0.2  \\
                  & 12 Aug 2017   &  0.46$\pm$0.03 &  $-$1.9$\pm$0.1   &    147$\pm$9        &                     & ~~~~~$>$$-$0.1  & ~~~~~$>$$-$0.1  & ~~\,2.1$\pm$0.2     & ~~\,2.9$\pm$0.2    & ~~~~~$>$$-$0.2  \\
                  & 28 Aug 2017   &  0.48$\pm$0.02 &  $-$1.8$\pm$0.1   &    153$\pm$6        &                     & ~~~~~$>$$-$0.2  & ~~~~~$>$$-$0.1  & ~~\,2.0$\pm$0.2     & ~~\,2.3$\pm$0.2    & ~~~~~$>$$-$0.1  \\
                  & 29 Aug 2017   &  0.45$\pm$0.03 &  $-$2.0$\pm$0.1   &    142$\pm$6        &                     & ~~~~~$>$$-$0.1  & ~~~~~$>$$-$0.1  & ~~\,2.0$\pm$0.2     & ~~\,2.7$\pm$0.2    & ~~~~~$>$$-$0.1  \\
                  & 17 Jul 2018   &  0.47$\pm$0.02 &  $-$1.4$\pm$0.1   &    156$\pm$10       &                     & ~~~~~$>$$-$0.1  & ~~~~~$>$$-$0.1  & ~~\,2.2$\pm$0.2     & ~~\,2.4$\pm$0.2    & ~~~~~$>$$-$0.2  \\
    \noalign{\smallskip}
    \hline 
   \end{tabular}
  $$
   \begin{list}{}{}
  \item[] {\bf Notes.} 
  \item[] D*: DENIS
  \item[] Columns 3, 4, 7, 8, 9, 10, and 11: Equivalent widths in \AA.
  \item[] No accretion rate is estimated for DENIS1538$-$1039, as we classify the source as a nonaccretor
    with the current data.
  \end{list}
  }
\end{table*}

\begin{figure}
   \centering
    \includegraphics[width=8cm,angle=0]{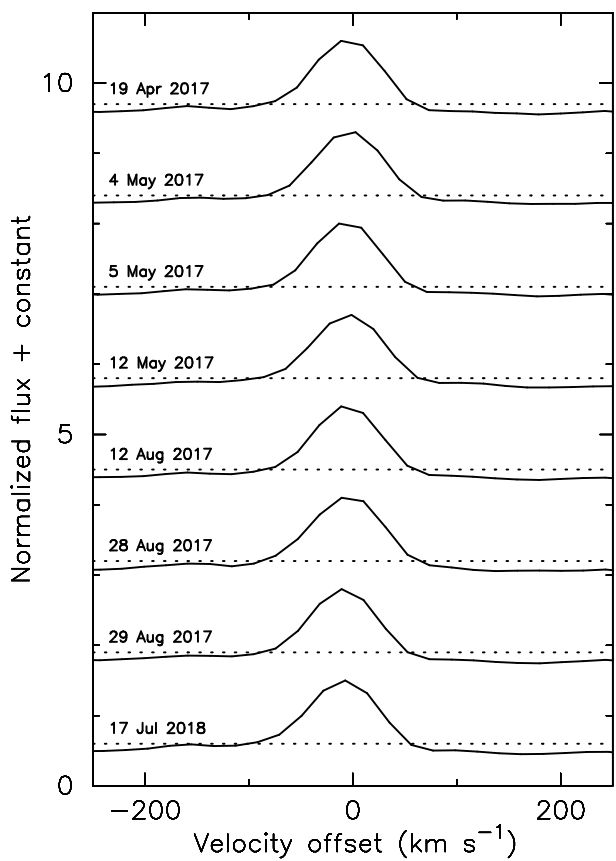}
    \caption{Same as Fig.~\ref{f7}, but for DENIS1538$-$1039.
}
\label{f9}
\end{figure}

For DENIS1538$-$1039, the $v_{10}$ values of all spectra are below the cutoff of 200~km~s$^{-1}$.
We therefore classify the M3.5 dwarf as nonaccretor. Figure~\ref{f9} shows the observed profiles of
H$\alpha$ emission in the dwarf.

\subsection{Accretion disk emission}
Based on the photometric data, the Gaia parallax of DENIS1538$-$1038,
and the BT-Settl models with solar metallicity as mentioned in Sect.~2,
our SED (see Fig.~\ref{f2}) yields a best-fit BT-Settl model of $T_{\star} = 2800\pm100$~K with $\log g = 3.5\pm0.5,$
and a blackbody with $T_{\rm disk} = 224$~K and $L_{\rm IR}/L_{\star} = 0.05$.
Our estimated $T_{\star}$ value is relevant for an M5.5$\pm$0.5 dwarf at 1~Myr
when compared to a well-measured value of $\sim$2700$\pm$100~K for an M6.5$\pm$0.5 BD
at a similar age \citep{stassun07}.

Our model SED poorly fits the $W2$ data point as seen in Fig.~\ref{f2}.
This is possibly due to the fact that
the nearby M3.5 dwarf DENIS1538$-$1039 is much brighter in the $W2$ band than
DENIS1538$-$1038 (see Fig.~\ref{f1} and Table~\ref{tab}). In addition,
the spatial resolution of the $W2$ image ($6.4''$) is poorer than that of $W1$ ($6.1''$).
Therefore, the flux measured for DENIS1538$-$1038 in $W2$ might be contaminated 
by DENIS1538$-$1039 much more than it is in $W1$ and $W3$.
As discussed in Sect.~2, the resolution of the $W4$ image could not spatially resolve the two M dwarfs.
Therefore, the measured $W4$ flux of DENIS1538$-$1038 that might include the flux from DENIS1538$-$1039
could affect our SED fitting. To subtract the $W4$-band flux from DENIS1538$-$1039,
we estimated the flux from the best-fit BT-Settl model
to the photometric data (except $W4$) of the source. We then reconstructed the SED of DENIS1538$-$1038.
The revised $W4$ flux yields a blackbody with $T_{\rm disk} = 300$~K
and $L_{\rm IR}/L_{\star} = 0.04$. 

\section{Summary}
In this paper, we present our detection of the accreting M5.5 dwarf, DENIS1538$-$1038,
based on its strong IR excesses in all four WISE bands.
Using the Gaia trigonometric paralax, we estimated an age of 1 Myr and a mass of 47~$M_{\rm J}$
for DENIS1538$-$1038. The lithium detection in DENIS1538$-$1038 also confirms its youth.
Based on H$\alpha$ emission line profiles, we estimated the accretion rate in DENIS1538$-$1038 in
the range of $10^{-7.9}-10^{-10.5}$~$M_{\odot}$~yr$^{-1}$.
These accretion rate values are generally lower than the typical values of
  $10^{-6}-10^{-8}$~$M_{\odot}$~yr$^{-1}$ for young low-mass stars (e.g., \citealt{hartigan})
  by about two orders of magnitude.
The observed H$\alpha$ variability indicates a sporadic accretion in the 1-Myr old BD.
Our discovery of the 1-Myr old BD that exhibits sporadic accretion with
low accretion rates supports a possible scenario for BD formation \citep{pb14} where
low-mass accretion rates at very early stages
(possibly with high outlfow mass-loss rate-to-mass-accretion-rate ratios)
prevent VLM cores from accreting enough gas to become stars, and 
thus these cores would end up as BDs.

\begin{acknowledgements} 
This research is funded by Vietnam National Foundation 
for Science and Technology Development (NAFOSTED) 
under grant number 103.99-2015.108.
This publication makes use of data products from the Wide-field Infrared Survey Explorer,
which is a joint project of the University of California, Los Angeles,
and the Jet Propulsion Laboratory/California Institute of Technology,
funded by the National Aeronautics and Space Administration.
This work has made use of data from the European Space Agency (ESA)
mission Gaia (https://www.cosmos.esa.int/gaia),
processed by the Gaia Data Processing and Analysis Consortium
(DPAC, https://www.cosmos.esa.int/web/gaia/dpac/consortium).
Funding for the DPAC has been provided by national institutions,
in particular the institutions participating in the Gaia Multilateral Agreement.
The DENIS project has been partly funded by the SCIENCE and the HCM plans of
the European Commission under grants CT920791 and CT940627.
It is supported by INSU, MEN and CNRS in France, by the State of Baden-W\"urttemberg 
in Germany, by DGICYT in Spain, by CNR in Italy, by FFwFBWF in Austria, by FAPESP in Brazil,
by OTKA grants F-4239 and F-013990 in Hungary, and by the ESO C\&EE grant A-04-046.
Jean Claude Renault from IAP was the Project manager.  Observations were  
carried out thanks to the contribution of numerous students and young 
scientists from all involved institutes, under the supervision of  P. Fouqu\'e,  
survey astronomer resident in Chile. 
This publication makes use of data products from the Two Micron All Sky Survey, which is a joint project of
the University of Massachusetts and the Infrared Processing and Analysis Center/California Institute of Technology,
funded by the National Aeronautics and Space Administration and the National Science Foundation.
This research has made use of the VizieR catalogue access tool, CDS,
Strasbourg, France. The original description of the VizieR service was
published in A\&AS 143, 23.
This research has made use of the SIMBAD database,
operated at CDS, Strasbourg, France.\\
We thank the referee, Elisabetta Rigliaco, for a useful report.
\end{acknowledgements}

\begin{appendix}
  \section{SED of late-M dwarfs}
\end{appendix}

\begin{figure*}
   \centering
     \includegraphics[width=17.0cm,angle=0]{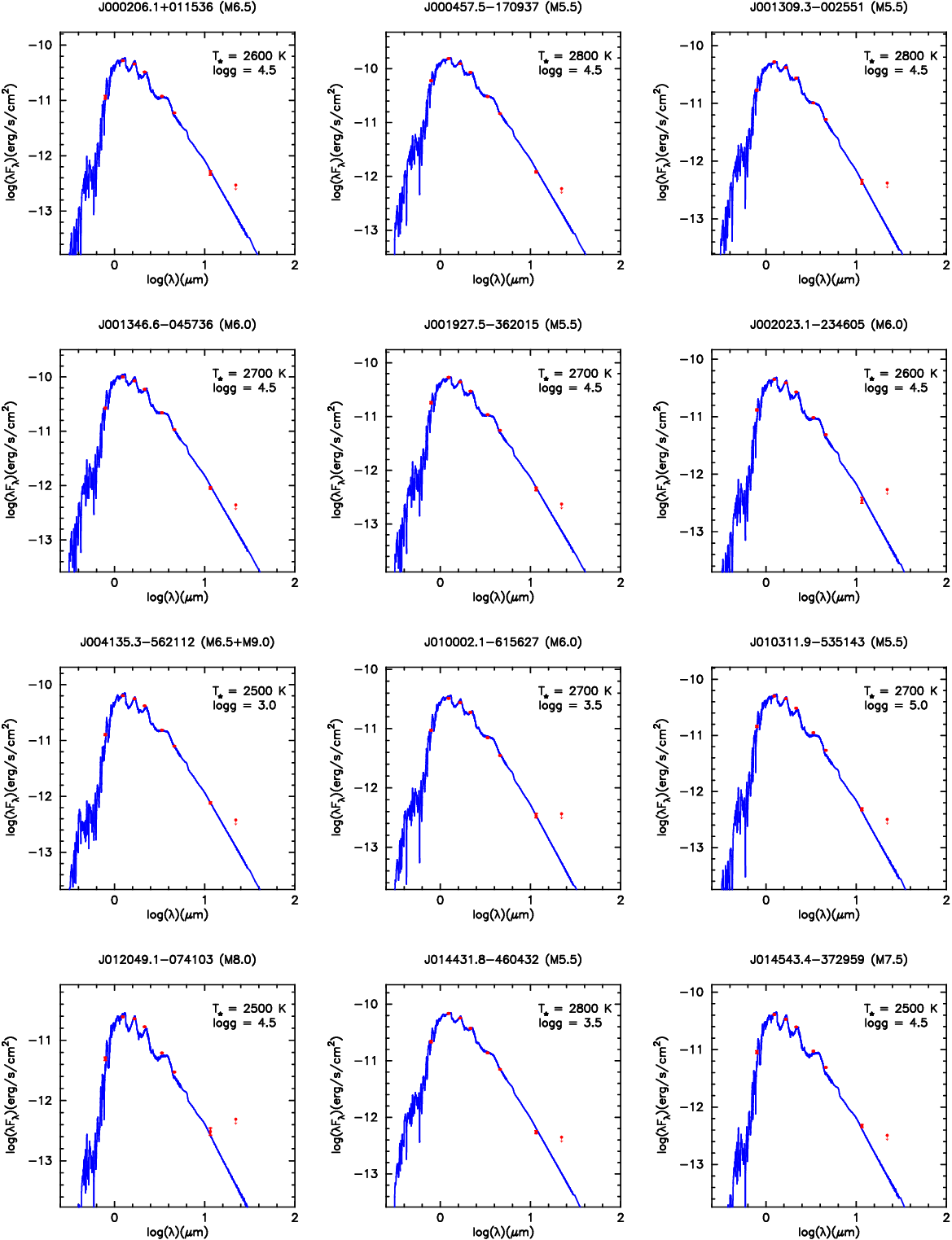}       
     \caption{
       Effective temperature $T_{\star}$ and surface gravity $\log g$ from the best-fit BT-Settl model (blue solid line)
       are shown. Spectral types are mainly taken from \citet{pb06a,crifo,pb17}. Except J004135.3$-$562112, which were taken 
       from \citet{reiners09}, J041048.0$-$125142 from \citet{pb06c}, 
       J084818.9$-$201911 from \citet{reid05}, J141159.9$-$413221 from \citet{pb17} and \citet{mont},
       J150416.1$-$235556 from \citet{reid03}, and J220622.7$-$204706 from \citet{crifo} and \citet{close}.
}
\label{f10a}
\end{figure*}

\begin{figure*}
  \setcounter{figure}{0}
   \centering
     \includegraphics[width=17.0cm,angle=0]{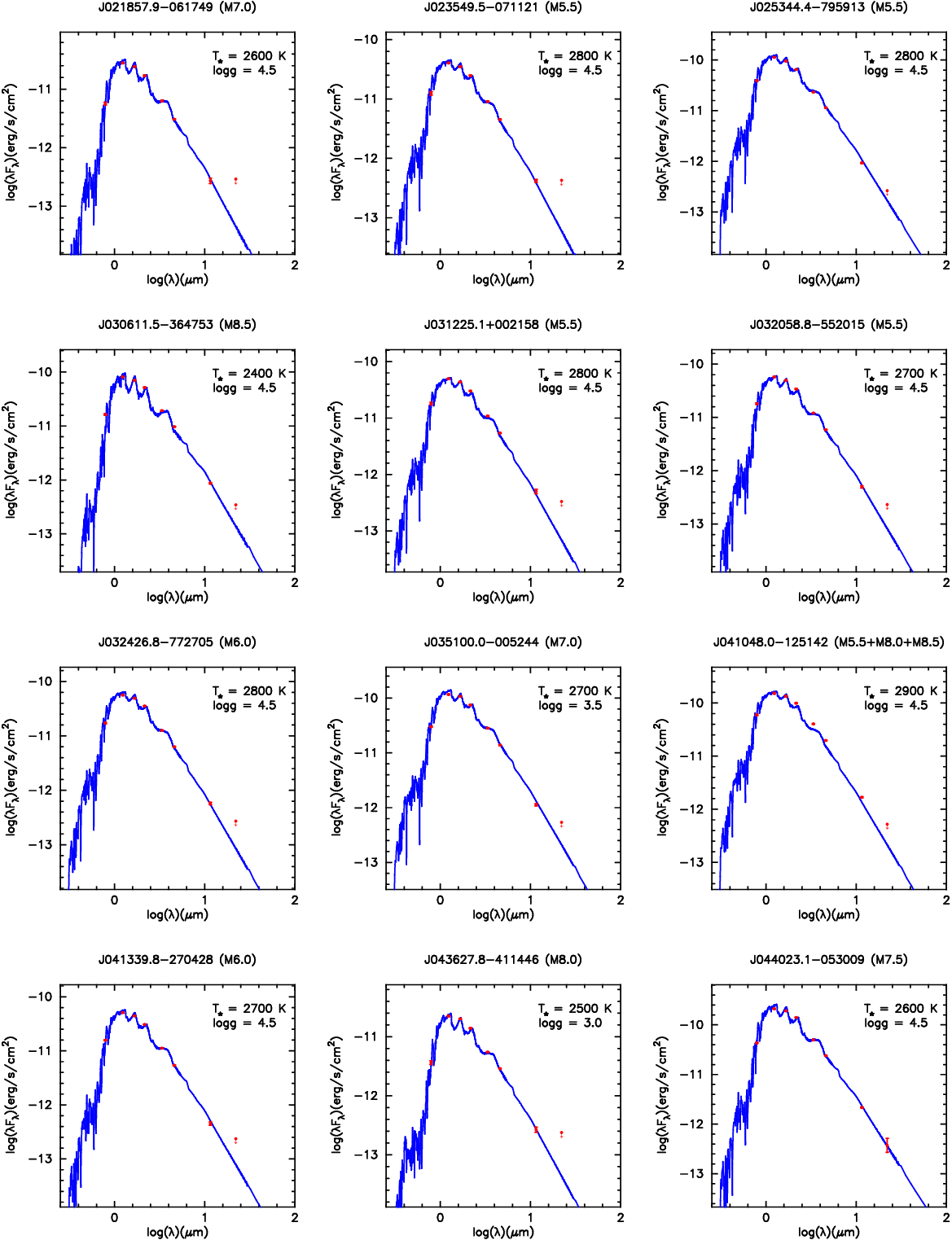}       
     \caption{continued.
}
\label{f10b}
\end{figure*}

\begin{figure*}
  \setcounter{figure}{0}
   \centering
     \includegraphics[width=17.0cm,angle=0]{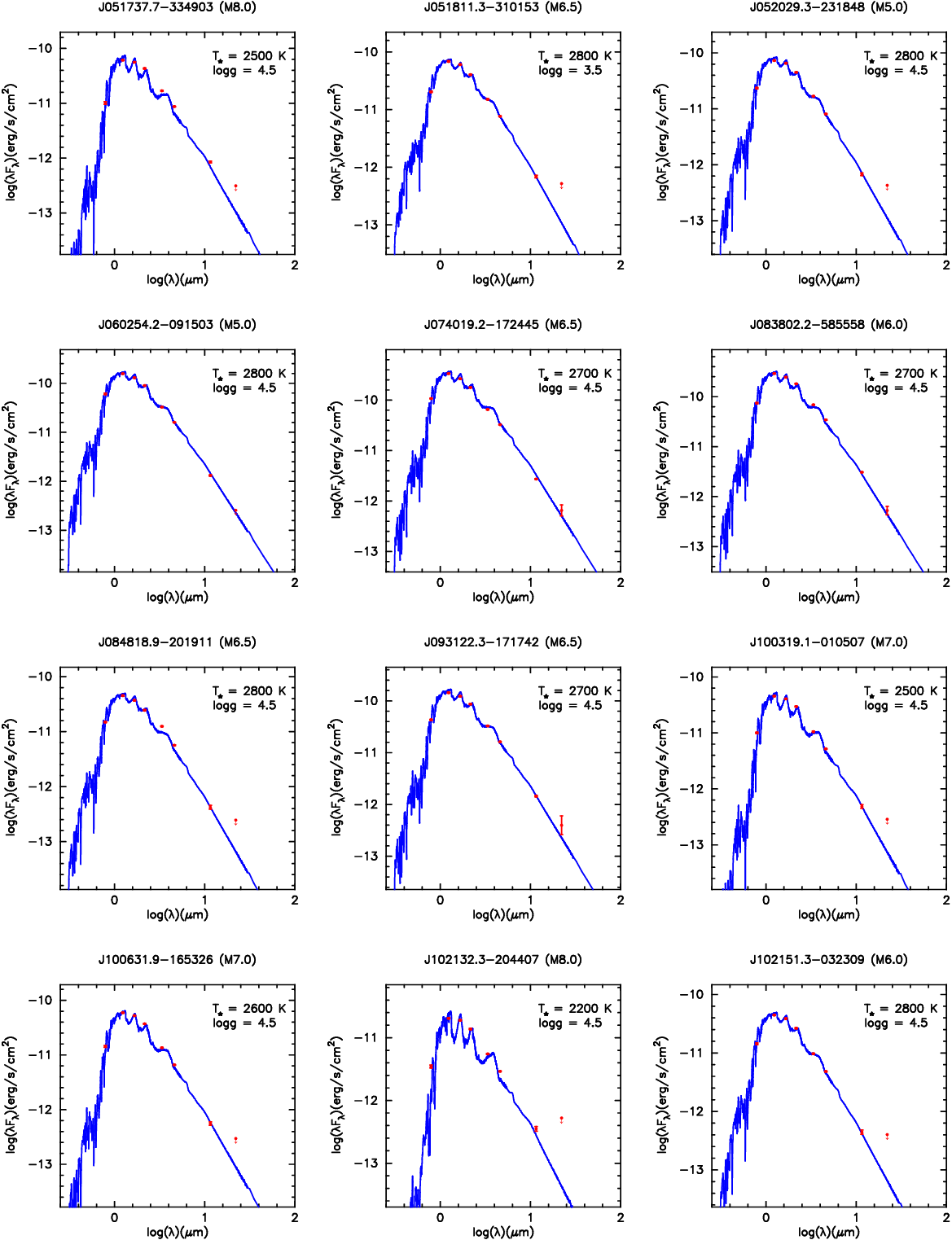}       
     \caption{continued.
}
\label{f10c}
\end{figure*}

\begin{figure*}
  \setcounter{figure}{0}
   \centering
     \includegraphics[width=17.0cm,angle=0]{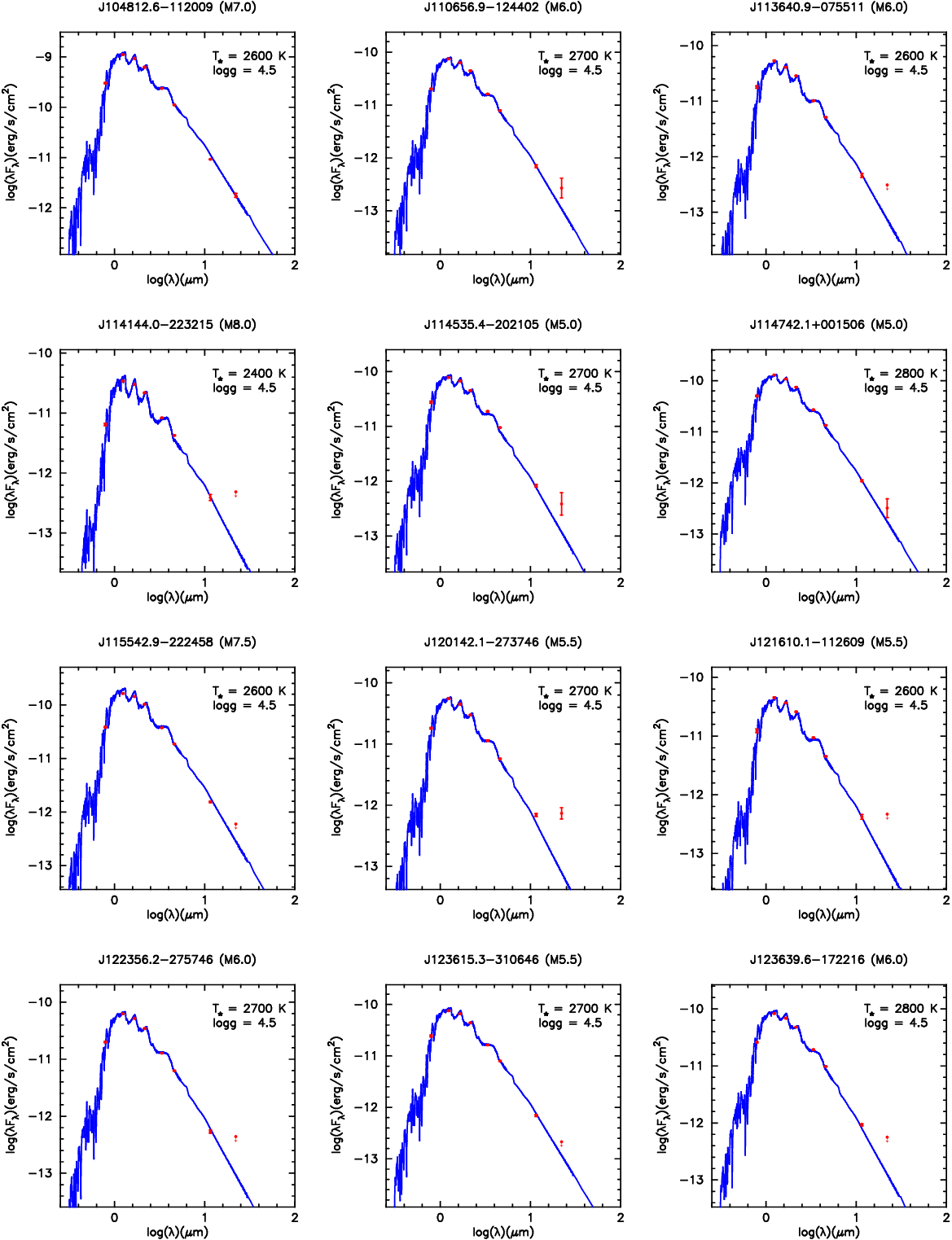}       
     \caption{continued.
}
\label{f10d}
\end{figure*}

\begin{figure*}
  \setcounter{figure}{0}
   \centering
     \includegraphics[width=17.0cm,angle=0]{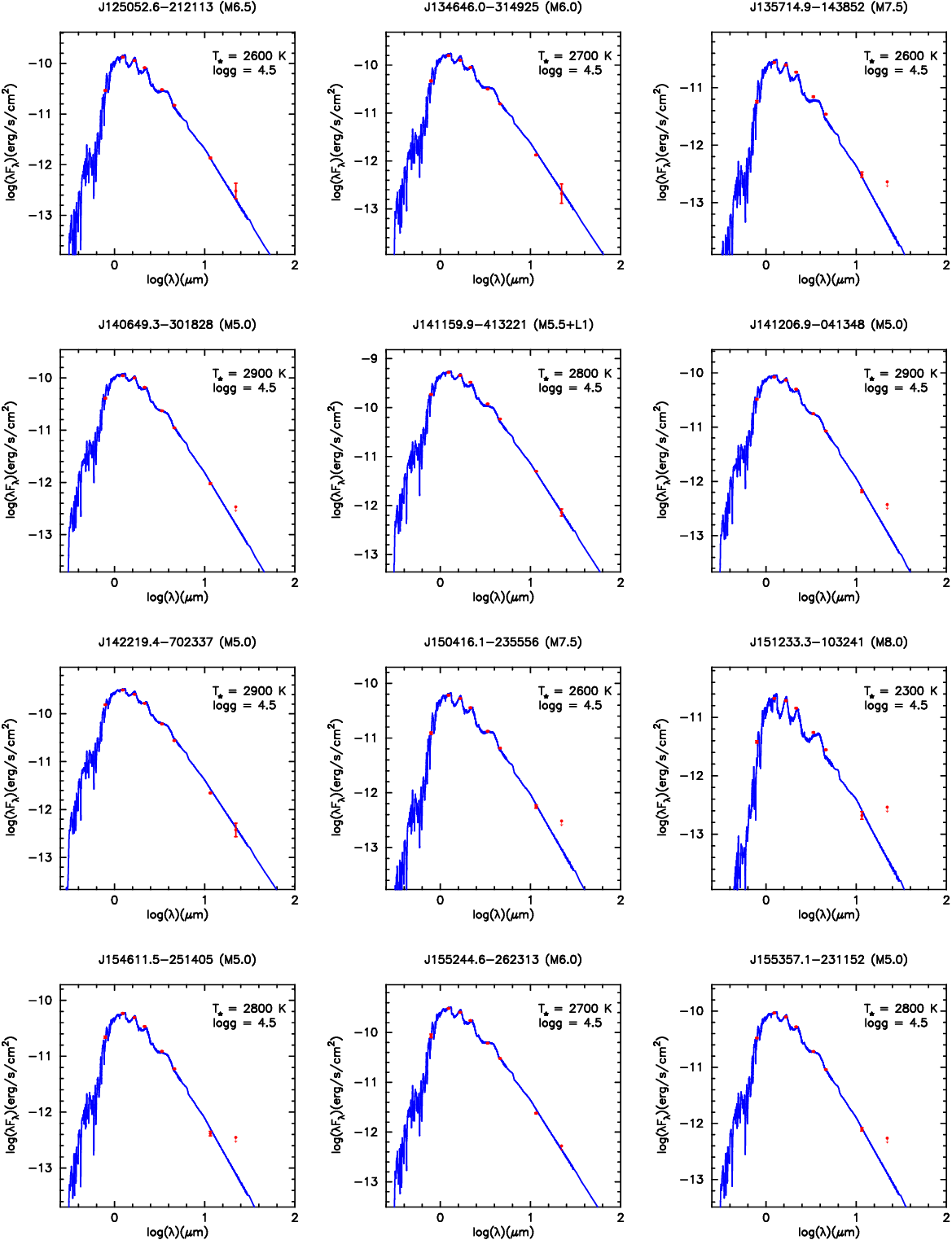}       
     \caption{continued.
}
\label{f10e}
\end{figure*}

\begin{figure*}
  \setcounter{figure}{0}
   \centering
     \includegraphics[width=17.0cm,angle=0]{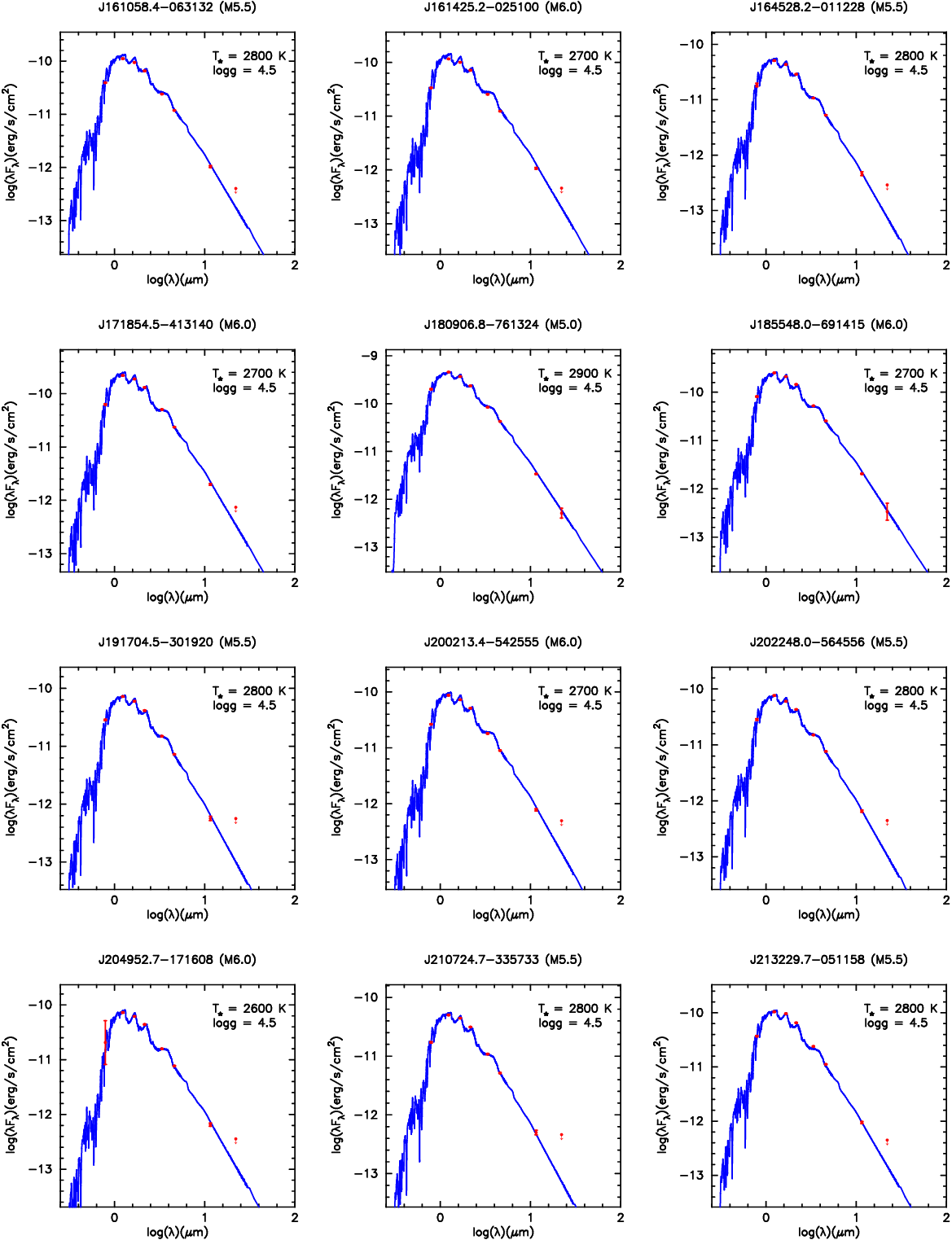}       
     \caption{continued.
}
\label{f10f}
\end{figure*}

\begin{figure*}
  \setcounter{figure}{0}
   \centering
     \includegraphics[width=17.0cm,angle=0]{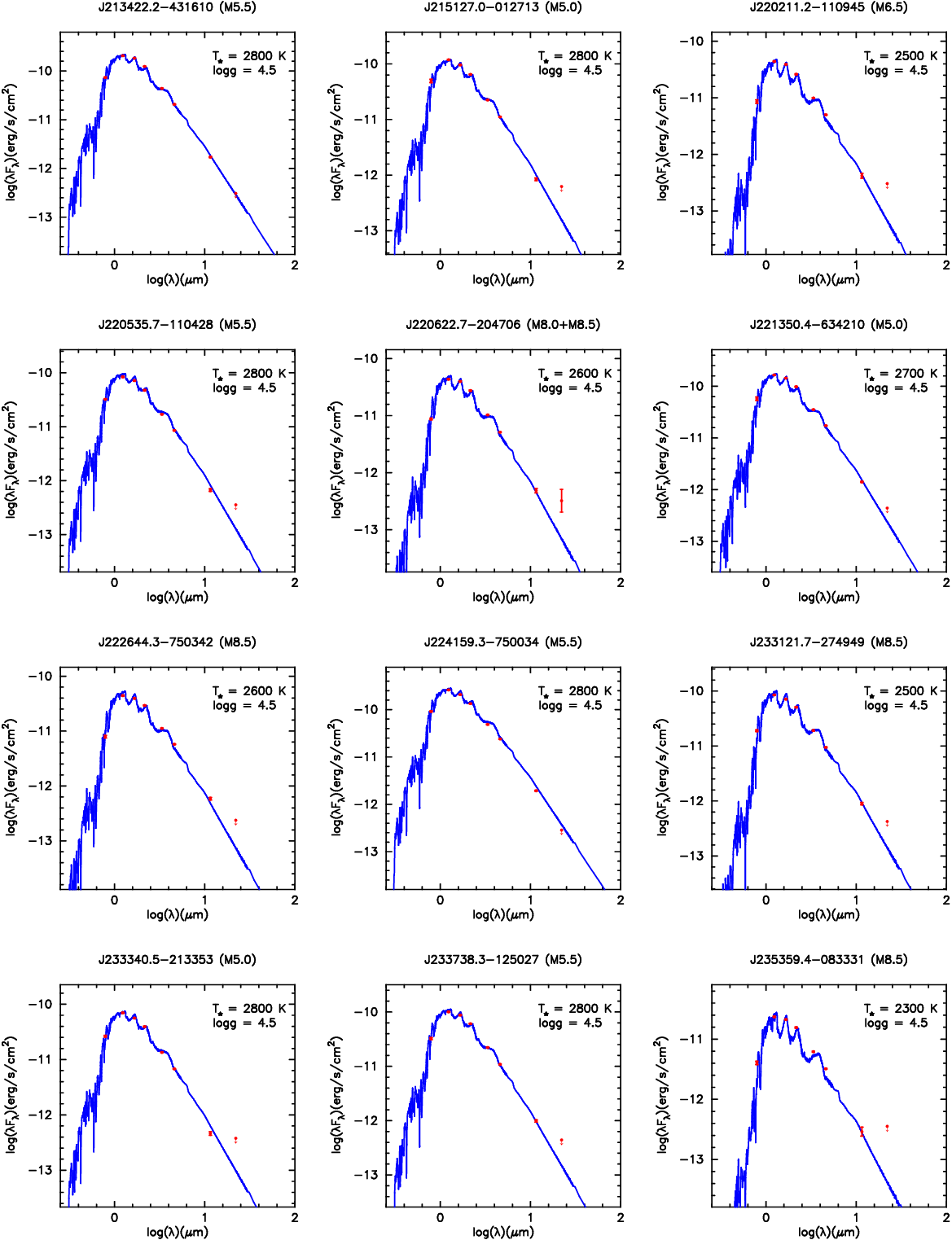}       
     \caption{continued.
}
\label{f10g}
\end{figure*}


\begin{thebibliography}{}

\bibitem[Allard et al.(2013)]{allard13}
  Allard, F., Homeier, D., Freytag, B., Schaffenberger, W., \& Rajpurohit, A. S. 2013, MmSAI, 24, 128

\bibitem[Avenhaus et al.(2012)]{aven}
  Avenhaus, H., Schmid, H. M., \& Meyer, M. R. 2012, A\&A, 548, A105

\bibitem[Basri(2000)]{basri00}
  Basri, G. 2000, ARA\&A, 38, 485

\bibitem[Bessell(1991)]{bessell91} 
  Bessell, M. S. 1991, AJ, 101, 662

\bibitem[Bessell(1999)]{bessell99} 
  Bessell, M. S. 1999, PASP, 111, 1426

\bibitem[Boucher et al.(2016)]{boucher16}
  Boucher, A., Lafreni\`ere, D., Gagn\'e, J., et al. 2016, ApJ, 832, 50

\bibitem[Cabrit et al.(1990)]{cabrit}
  Cabrit, S., Suzan, E., Strom, S. E., \& Strom, K. M. 1990, ApJ, 354, 687 

\bibitem[Close et al.(2002)]{close}
   Close, L. M., Siegler, N., Potter, D., Brandner, W., \& Liebert, J. 2002, ApJ, 567, L53

\bibitem[Cohen et al.(2003)]{cohen}
  Cohen, M., Wheaton, WM. A., \&  Megeath, S. T. 2003, AJ, 126, 1090 

\bibitem[Comer\'on et al.(2003)]{com}
  Comer\'on, F., Fern\'andez, M., Baraffe, I., Neuh\"auser, R., \& Kaas, A. A. 2003, A\&A, 406, 1001

\bibitem[Comer\'on et al.(2010)]{comeron10}
  Comer\'on, F., Testi, L., \& Natta, A. 2010, A\&A, 522, 47

\bibitem[Crifo et al.(2005)]{crifo}
  Crifo, F., Phan-Bao, N., Delfosse, X., et al. 2005, A\&A, 441, 653

\bibitem[Cutri et al.(2014)]{cutri14}
  Cutri, R. M., et al. 2014, VizieR On-line Data Catalog, 2311, 0

\bibitem[Dang-Duc et al.(2016)]{cuong}
  Dang-Duc, C., Phan-Bao, N., \&  Dao-Van, D. T. 2016, A\&A, 588, L2

\bibitem[Dopita et al.(2007)]{dopita}
  Dopita, M., Hart, J., McGregor, P., et al. 2007, Ap\&SS, 310, 255

\bibitem[Flaherty et al.(2019)]{flaherty}
  Flaherty, K., Hughes, A. M., Mamajek, E. E., \& Murphy, S. J. 2019, ApJ, 872, 92 

\bibitem[Fouqu\'e et al.(2000)]{fouque}
 Fouqu\'e, P., Chevallier, L., Cohen, M., et al. 2000, A\&ASS, 141, 313

\bibitem[Gagn\'e et al.(2018)]{gagne18}
  Gagn\'e, J., Mamajek, E. E., Malo, L., et al. 2018, ApJ, 856, 23

\bibitem[Hartigan et al.(1995)]{hartigan} 
 Hartigan, P., Edwards, S. \& Ghandour, L. 1995, ApJ, 452, 736

\bibitem[Herczeg \& Hillenbrand (2008)]{herc} 
 Herczeg, G. J., \& Hillenbrand, L. A. 2008, ApJ, 681, 594

\bibitem[Jayawardhana et al.(2003)]{jay03} 
  Jayawardhana, R., Mohanty, S. \& Basri, G. 2003, ApJ, 592, 282

\bibitem[Jayawardhana et al.(2006)]{jay06} 
  Jayawardhana, R., Coffey, J., Scholz, A., Brandeker, A., \& Van Kerkwijk, M. H. 2006, ApJ, 648, 1206

 \bibitem[Lawson et al.(2004)]{lawson}
   Lawson, W. A., Lyo, A-Ran, \& Muzerolle, J. 2004, MNRAS, 351, L39

 \bibitem[Luhman et al.(2007)]{luhman07}
   Luhman, K. L., Joergens, V., Lada, C., et al. 2007, in Protostars and Planets V,
   ed. B. Reipurth, D. Jewitt, \& K. Keil (Tucson, AZ: Univ. Arizona Press), 443

 \bibitem[Luhman et al.(2018)]{luhman18}
 Luhman, K. L., Herrmann, K. A., Mamajek, E. E., Esplin, T. L., \& Pecault, M. J. 2018, AJ, 156, 76

\bibitem[Manara et al.(2015)]{manara} 
  Manara, C. F.,  Testi, L., Natta, A., \& Alcal\'a, J. M. 2015, A\&A, 579, A66

\bibitem[Mart\'{\i}n et al.(2010)]{martin10} 
  Mart\'{\i}n, E. L., Phan-Bao, N., Bessell, M. S., et al. 2010, A\&A, 517, 53 

\bibitem[Montagnier et al.(2006)]{mont} 
  Montagnier, G., S\'egransan, D., Beuzit, J.-L. 2006, A\&A, 460, L19

\bibitem[Murphy et al.(2011)]{murphy11}
  Murphy, S. J., Lawson, W. A., Bessell, M. S., \& Bayliss, D. D. J. 2011, MNRAS, 411, L51

\bibitem[Murphy \& Lawson(2015)]{murphy15a}
  Murphy, S. J., \& Lawson, W. A. 2015, MNRAS, 447, 1267

\bibitem[Murphy et al.(2015)]{murphy15}
  Murphy, S. J., Lawson, W. A., \& Bento, J. 2015, MNRAS, 453, 2220

\bibitem[Murphy et al.(2018)]{murphy18}
  Murphy, S. J., Mamajek, E. E., \& Bell, C. P. M. 2018, MNRAS, 476, 3290

\bibitem[Muzerolle et al.(2003)]{muz03}
  Muzerolle, J., Hillenbrand, L., Calvet, N., Brice\~{n}o, C., \& Hartmann, L. 2003, ApJ, 592, 266

\bibitem[Muzerolle et al.(2005)]{muz05}
  Muzerolle, J., Luhman, K. L., Brice\~{n}o, C., Hartmann, L., \& Calvet, N. 2005, ApJ, 625, 906

\bibitem[Natta et al.(2004)]{natta} 
  Natta, A., Testi, L., Muzerolle, J., et al. 2004, A\&A, 424, 603

\bibitem[Pavlenko et al.(1995)]{pav}
  Pavlenko, Y. V., Rebolo, R., Mart{\'{\i}}n E. L., \& Garc{\'{\i}}a L{\'o}pez, R. J. 
  1995, A\&A, 303, 807

\bibitem[Pecaut \& Mamajek(2013)]{pecaut}
  Pecaut, M. J., \& Mamajek, E. E. 2013, ApJS, 208, 9

 \bibitem[Phan-Bao et al.(2001)]{pb01}
  Phan-Bao, N., Guibert, J., Crifo, F., et al. 2001, A\&A, 380, 590

\bibitem[Phan-Bao et al.(2003)]{pb03}
  Phan-Bao, N., Crifo, F., Delfosse, X., et al. 2003, A\&A, 401, 959

\bibitem[Phan-Bao \& Bessell(2006)]{pb06a}
  Phan-Bao, N., \& Bessell M. S. 2006, A\&A, 446, 515

\bibitem[Phan-Bao et al.(2006)]{pb06c}
  Phan-Bao, N., Forveille, T., Mart{\'{\i}}n, E. L., \& Delfosse, X. 2006, ApJ, 645, L153

\bibitem[Phan-Bao et al.(2014)]{pb14}
  Phan-Bao, N., Lee, C.-F., Ho, P. T. P., Dang-Duc, C., \& Li, D. 2014, ApJ, 795, 70

\bibitem[Phan-Bao et al.(2017)]{pb17}
  Phan-Bao, N., Bessell, M. S., Nguyen-Thanh, D., et al. 2017, A\&A, 600, A19

\bibitem[Reid et al.(2003)]{reid03} 
  Reid, I. N., Cruz, K. L., Allen, P., et al. 2003, AJ, 126, 3007

\bibitem[Reid \& Gizis(2005)]{reid05} 
  Reid, I. N., \& Gizis, J. E. 2005, PASP, 117, 676

\bibitem[Reiners(2009)]{reiners09} 
  Reiners, A. 2009, ApJ, 702, L119

\bibitem[Reiners et al.(2010)]{reiners10} 
  Reiners, A., Seifahrt, A., \& Dreizler, S. 2010, A\&A, 513, L9

\bibitem[Rigliaco et al.(2012)]{rigliaco} 
  Rigliaco, E., Natta, A., Testi, L., Randich, S., Alcal\`a, J. M., et al. 2012, A\&A, 548, A56 

\bibitem[Riedel et al.(2019)]{rie19}
  Riedel, A. R., DiTomasso, V., Rice, E. L., et al. 2019, AJ, 157, 247

\bibitem[Rizzuto et al.(2015)]{riz15}
  Rizzuto, A. C., Ireland, M. J., \& Kraus, A. L. 2015, MNRAS, 448, 2737

\bibitem[Santamar\'{\i}a-Miranda et al.(2018)]{sm18}
  Santamar\'{\i}a-Miranda, A., C\'aceres, C., Schreiber, M. R., et al. 2018, MNRAS, 475, 2994

\bibitem[Scholz \& Jayawardhana(2006)]{scholz} 
  Scholz, A., \& Jayawardhana, R. 2006, ApJ, 638, 1056 

\bibitem[Shortridge et al.(2004)]{shortridge}
 Shortridge, K., Meyerdierks, H., Currie, M., et al. 2004, Starlink User Note 86

\bibitem[Silverberg et al.(2016)]{sil16} 
  Silverberg, S. M., Kuchner, M. J., Wisniewski, J. P., et al. 2016, ApJ, 830, L28

\bibitem[Stassun et al.(2007)]{stassun07} 
  Stassun, K. G., Mathieu, R. D., \& Valenti, J. A. 2007, ApJ, 664, 1154 

\bibitem[White \& Basri(2003)]{white} 
  White, R. J., \& Basri, G. 2003, ApJ, 582, 1109

\bibitem[Wright et al.(2010)]{wright} 
  Wright, E. L., Eisenhardt, P. R. M., Mainzer, A. K., et al. 2010, AJ, 140, 1868
  

\end{thebibliography}
\end{document}